\documentclass[]{spie}
\usepackage{amsmath,amsfonts,amssymb}
\usepackage{graphicx}
\usepackage[colorlinks=true, allcolors=blue]{hyperref}
\usepackage{booktabs}
\usepackage{float}
\usepackage{svg}
\usepackage{enumitem}

\title{He`e-Lab: A modular testbed for astrophotonics and wavefront sensing development}

\author[a,b,c]{S. Vievard}
\author[a,b]{C. Hamner}
\author[b]{N. Skaf}
\author[d]{D. Hively}
\author[e]{M. Bottom}
\author[b]{M. Chun}
\author[b]{C. Baranec}
\author[c,f,g]{O. Guyon}
\author[c,g]{J. Lozi}
\author[b]{A. Walk}
\author[b]{J. Corn}
\author[a,b]{B. Allen}
\author[h]{E. Huby}
\author[h]{S. Lacour}
\author[i]{G. Martin}
\author[i]{M. Lallement}
\author[c,j]{V. Deo}
\affil[a]{Space Science and Engineering Initiative, College of Engineering, University of Hawai‘i, Hilo, HI 96720, USA}
\affil[b]{Institute for Astronomy, University of Hawaii at Manoa, Hilo, HI 96720 USA}
\affil[c]{Subaru Telescope, National Astronomical Observatory of Japan, Hilo, HI, USA}
\affil[d]{University of Hawai‘i in Hilo, Hilo, HI 96720, USA}
\affil[e]{Department of Astronomy, University of California, Berkeley CA 94720, USA}
\affil[f]{Steward Observatory, University of Arizona, Tucson, AZ 85721, USA}
\affil[g]{Astrobiology Center, Mitaka, Tokyo 181-8588, Japan}
\affil[h]{LESIA, Observatoire de Paris, Université PSL, Meudon, France}
\affil[i]{Univ. Grenoble Alpes, CNRS, IPAG, 38400 Saint-Martin-d'Hères, France}
\affil[j]{Optical Sharpeners, Manosque, France}

\authorinfo{Further author information: (Send correspondence to S.V.)\\S.V.: E-mail: vievard@hawaii.edu}

\begin{document}
\maketitle

\begin{abstract}
Advanced astronomical instrumentation requires accessible, reconfigurable platforms to validate novel technologies and algorithms before on-sky deployment. We present the design, architecture, and alignment validation of the Hawaii Experimental Engineering Lab (He`e-Lab), a state-of-the-art modular testbed dedicated to two complementary research tracks: (A) the integration and characterization of astrophotonics components within a real-time computing loop, and (B) the development of advanced wavefront sensing and control (WFS\&C) algorithms. The testbed features a broadband supercontinuum source (500~nm to 2~$\mu$m), a high-order 1k-actuator Boston Micromachines deformable mirror, and a 37-segment hexagonal mirror assembly providing piston-tip-tilt control to emulate segmented apertures like Keck and JWST. Downstream capabilities include a HASO 126 Shack-Hartmann sensor, a real-time computing environment driven by the CACAO package, and a modular injection platform coupled to a visible-wavelength spectrograph ($R \sim 3,000$). We report on the successful system alignment and outline the roadmap for upcoming adaptive optics and photonic device validation frameworks.
\end{abstract}


\section{INTRODUCTION}
\label{sec:intro}
The current and next generation of high-contrast imaging and spectroscopic instruments for ground- and space-based astronomy demands precise control over wavefront aberrations and high-efficiency coupling into sub-diffraction-limited devices~\cite{lancaster2018decadal,jovanovic20232023}. Validating advanced instrumentation technologies—such as photonics, novel wavefront sensors, and multi-segment cophasing control schemes—carries operational risk, high cost, and lost observing time if executed directly on-sky. While on-sky development platforms like the Subaru Coronagraphic Extreme Adaptive Optics (SCExAO~\cite{jovanovic_subaru_2015}) platform have proven vital for maturing cutting-edge technologies in real telescope environments, their dual role as both science instruments and development platforms creates operational constraints. Located at high elevation and integrated directly within the telescope infrastructure, access to these systems is naturally limited, and hardware or software interventions carry elevated operational risks.

Consequently, there is a need for lower-risk, sea-level laboratory testbeds that mirror (as close as possible) the software and hardware environment of on-sky platforms. The Hawaii Experimental Engineering Lab (He`e-Lab) has been established to meet these demands through a highly modular and open-access hardware architecture. By offering a dual-track framework spanning active wavefront control and astrophotonics injection, He`e-Lab bridges the gap between laboratory bench concepts and eventual deployment on major facilities like the Subaru Telescope and the W. M. Keck Observatory. 

To achieve this, the testbed is designed around three core operational pillars: 
\begin{itemize}[nosep]
    \item \textbf{Risk Mitigation and Pre-Deployment Testing} : by mimicking both the hardware configuration and the real-time control software architecture (Compute and Control for Adaptive Optics, CACAO~\cite{guyon2020adaptive}) of on-sky instruments like SCExAO, the He`e-Lab provides visiting engineering teams with an accessible sea-level facility. Visiting teams can integrate novel hardware, debug interfaces, and optimize control loops in a controlled environment prior to high-altitude deployment.
    \item \textbf{Technology Development and Validation} : The testbed enables characterization of novel astrophotonic components—such as photonic lanterns and integrated photonic devices—alongside advanced Wavefront Sensing and Control (WFS\&C) strategies, including focal-plane WFS and segmented mirror cophasing.
    \item \textbf{Hands-on Education and Workforce Development :} Situated outside the operational constraints of an active telescope, the He`e-Lab serves as a low-risk training sandbox for undergraduate and graduate students. Early-career researchers gain direct experience with facility-grade hardware, alignment protocols, and real-time AO control systems without risking valuable telescope observing time.
\end{itemize}

The name of the testbed carries deep cultural significance grounded in indigenous Hawaiian knowledge. In 'Ōlelo Hawai'i (the Hawaiian language), \textit{he'e} translates to the octopus~\cite{pukui1986hawaiian}, a creature renowned for its extraordinary adaptability, intelligence, and fluid dexterity. In Hawaiian tradition, the \textit{he'e} is recognized as a \textit{kinolau} - the physical manifestation or bodily form - of Kanaloa, one of the primary gods of the ocean, navigation, and deep-sea realms~\cite{beckwith1970hawaiian, malo1951hawaiian}. Incorporating \textit{He'e} into the laboratory's identity reflects the testbed's core design philosophy: a highly flexible, multi-limbed platform capable of dynamically adapting to diverse technical challenges across adaptive optics, wavefront sensing, and astrophotonics.

In this paper, we describe the design, hardware architecture, software integration, and initial capabilities of the He`e-Lab testbed.

\section{TESTBED ARCHITECTURE AND HARDWARE SPECIFICATIONS}
\label{sec:design}
\begin{figure}
    \centering
    \includesvg[width=1\linewidth]{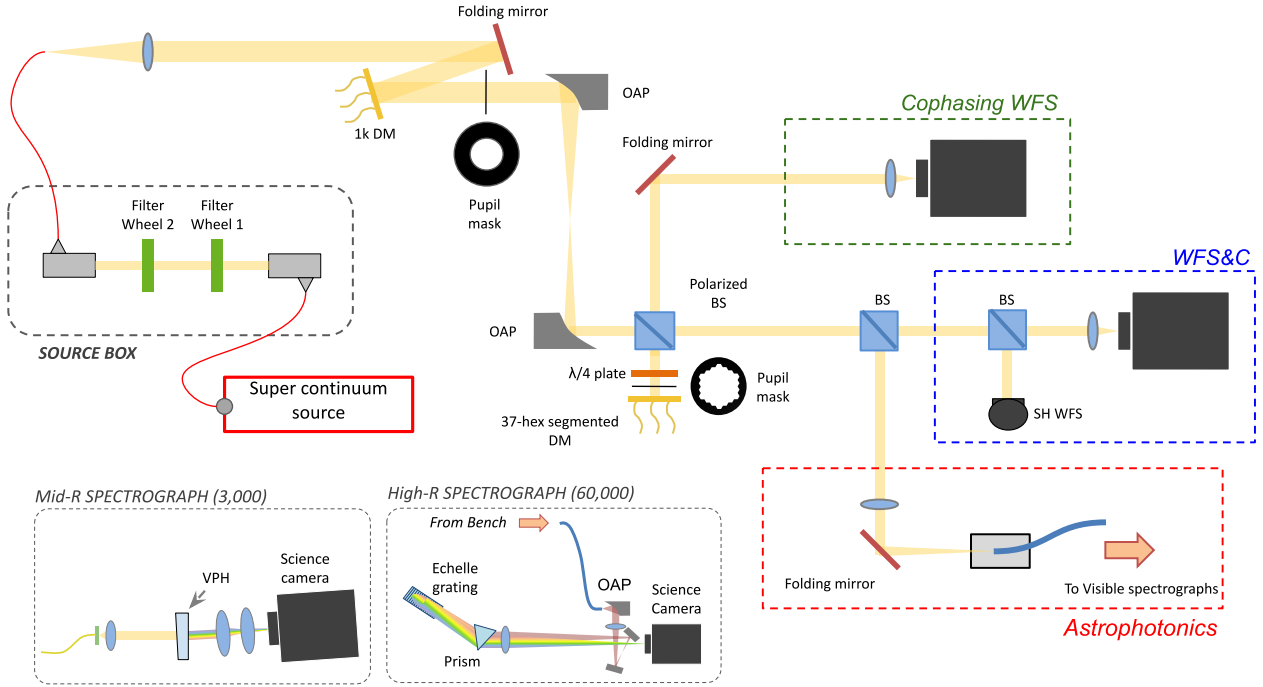}
    \caption{Optical layout of the He`e-Lab testbed. Colored dashed regions isolate the Source Box (gray), Cophasing WFS path (green), primary WFS\&C subsystem (blue), and the downstream Astrophotonics injection module (red)}
    \label{fig:placeholder}
\end{figure}

The He`e-Lab aims to facilitate collaborative research and technology development through a highly modular testbed designed around two complementary objectives:
\begin{enumerate}[nosep]
    \item Integration, characterization, and validation of \textbf{astrophotonic components} within a real-time computing environment prior to telescope deployment.
    \item Development and validation of \textbf{advanced wavefront sensing and control (WFS\&C)} algorithms.
\end{enumerate}
The principal hardware components and system parameters are summarized in Table~\ref{tab:specs}. The optical layout of the testbed is shown in Figure~\ref{fig:placeholder}.

\begin{table}[htbp]
\caption{He`e-Lab system specifications and core hardware components.}
\label{tab:specs}
\centering
\begin{tabular}{ll}
\hline
\textbf{Subsystem / Parameter} & \textbf{Technical Specification} \\ \hline
Illumination Source & NKT COMPACT Supercontinuum ($500\,\text{nm}$--$2\,\mu\text{m}$) \\
High-Order WFC & Boston Micromachines (BMC) 1k DM ($952$ active actuators, $1.5\,\mu\text{m}$ stroke) \\
Segmented Aperture Simulator & 37-Hexagonal Segmented Mirror (3 actuators/segment: Piston-Tip-Tilt) \\
High-Order Wavefront Sensor & HASO 126 Shack-Hartmann WFS \\
Mid-Resolution Spectroscopy & VPH-based Spectrograph ($R \sim 3{,}000$, $600$--$800\,\text{nm}$) \\
Real-Time Software Control & CACAO (Compute and Control for Adaptive Optics) Package \\
\hline
\end{tabular}
\end{table}

\subsection{Source Module and Beam Conditioning}
We use a NKT COMPACT supercontinuum fiber laser source, providing a spectral continuum spanning 500~nm to 2~$\mu$m. Light exiting the fiber is collimated using a Thorlabs RC12FC-P01 reflective collimator equipped with a protected silver coating to guarantee high throughput and eliminate chromatic aberrations across the full operating band. Spectral selection and intensity attenuation are managed via a dual-stage filter wheel configuration:
\begin{itemize}[nosep]
    \item \textbf{Filter Slider 1 (Visible ND):} Populated with Neutral Density (ND 4.0, ND 2.0, ND 1.0) filters and an open slot for coarse beam intensity adjustments.
    \item \textbf{Filter Slider 2 (IR \& Spectral):} Equipped with IR 4.0 and IR 2.0 filters for near-infrared attenuation, an open slot, and a narrow-band 650~nm filter for monochromatic visible alignment and optimization.
\end{itemize}
A secondary matched RC12FC-P01 reflective collimator delivers the conditioned, stable parallel beam downstream into the primary adaptive optics path.

\subsection{Wavefront Manipulation Subsystems}
The testbed integrates two distinct active deformable mirrors dedicated to high-order aberration correction, turbulence simulation, and segmented cophasing research:
\begin{itemize}[nosep]
    \item \textbf{1k Deformable Mirror:} A continuous-surface Boston Micromachines Corporation (BMC) deformable mirror featuring 952 active actuators across a 34-actuator wide clear aperture, providing up to $1.5\,\mu\text{m}$ of physical actuator stroke. This device enables high-spatial-frequency phase correction, static aberration correction, or atmospheric turbulence simulation.
    \item \textbf{37-Segment Hexagonal Mirror:} A dedicated 37-segment hexagonal segmented mirror array designed to mimic multi-segment architectures such as the W. M. Keck Observatory and the James Webb Space Telescope (JWST). Each individual hexagonal segment is driven by three independent actuators providing precise Piston, Tip, and Tilt (PTT) degrees of freedom.
\end{itemize}

\subsection{Optical Train and Polarization-Engineered Cophasing Arm}
The optical beam propagates downstream from the light source delivery system and encounters the continuous BMC 1k DM for primary high-order wavefront correction. Immediately following the 1k DM, the beam reaches a Polarizing Beam Splitter (PBS) cube that efficiently divides the polarization states:
\begin{itemize}[nosep]
    \item \textbf{Transmitted Beam (P-polarization):} Directs the primary science/diagnostic light toward the downstream wavefront sensing and control (WFS\&C) sensors and the astrophotonic injection platform.
    \item \textbf{Reflected Beam (S-polarization):} Enters a double-pass cophasing arm designed to maximize photon throughput. The reflected light passes through a Quarter-Wave Plate ($\text{QWP}$), reflects off the 37-segment PTT mirror (which includes an adjacent hexagonal pupil mask to match segment geometry), and passes back through the $\text{QWP}$. 
\end{itemize}
This dual pass through the $\text{QWP}$ rotates the linear polarization state by $90^\circ$ (transitioning $S \rightarrow P$), allowing the returning beam to be fully transmitted back through the PBS cube directly toward the cophasing camera without splitting loss.

\section{INSTRUMENTATION AND CONTROL FRAMEWORK}
\label{sec:instrumentation}

\subsection{Wavefront Sensing Configurations}
The testbed supports multiple concurrent sensing architectures. High-order residual phase tracking is performed by a HASO 126 Shack-Hartmann wavefront sensor positioned in a dedicated adaptive optics path (WFSC\&C in blue on the optical layout). In parallel, a secondary path is reserved for segmented cophasing sensing developments (e.g., phase diversity and focal plane algorithms) utilizing a dedicated Cophasing Wavefront Sensor camera (Cophasing WFS in green on the optical layout).

\subsection{Modular Astrophotonics Injection Platform}
The astrophotonics track branches from the main optical relay via an injection module that feeds into a mid-resolution ($R \sim 3{,}000$) Volume Phase Holographic (VPH) spectrograph~\cite{vievard_spectroscopy_2024} optimized for visible wavelengths ($600$--$800\,\text{nm}$). To maximize adaptability, the fiber injection module features a swappable interface mounting plate. While currently configured for Single-Mode Fibers (SMF), Multi-Mode Fibers (MMF), and Photonic Lantern devices, the plate can be easily exchanged to accommodate 3D-written integrated photonic chips. 

Coupling optimization is handled dynamically via a high-precision 3-axis ($XYZ$) mechanical translation stage equipped with remotely operable Zaber motorized actuators on the transverse ($XY$) axes, combined with active phase profile optimizations running upstream on the continuous 1k DM.

\subsection{Software Architecture and Control Environment}
The software architecture is built around a two-tiered framework designed for low-latency hardware control and high-speed telemetry streaming. At the hardware abstraction level, device interfaces—including the Zaber motorized stages, cameras, and filter wheels—are driven using \texttt{pyMilk}, a Python library that utilizes shared-memory structures for fast inter-process communication. 

This shared-memory interface integrates naturally with the Compute and Control for Adaptive Optics (\texttt{CACAO}) software environment, which handles high-level hardware synchronization, real-time wavefront reconstruction, and loop execution. By passing telemetry seamlessly between \texttt{pyMilk} and \texttt{CACAO}, the integrated framework enables deterministic, low-latency operations capable of closing active feedback loops simultaneously across the continuous 1k DM grid and the 37-segment PTT mirror assembly.

\section{COMMISSIONING STATUS AND ALIGNMENT VALIDATION}
\label{sec:commissioning}

\subsection{Full Bench Alignment and Wavefront Sensing Integration}
\begin{figure}[htbp]
    \centering
    \includegraphics[width=\linewidth]{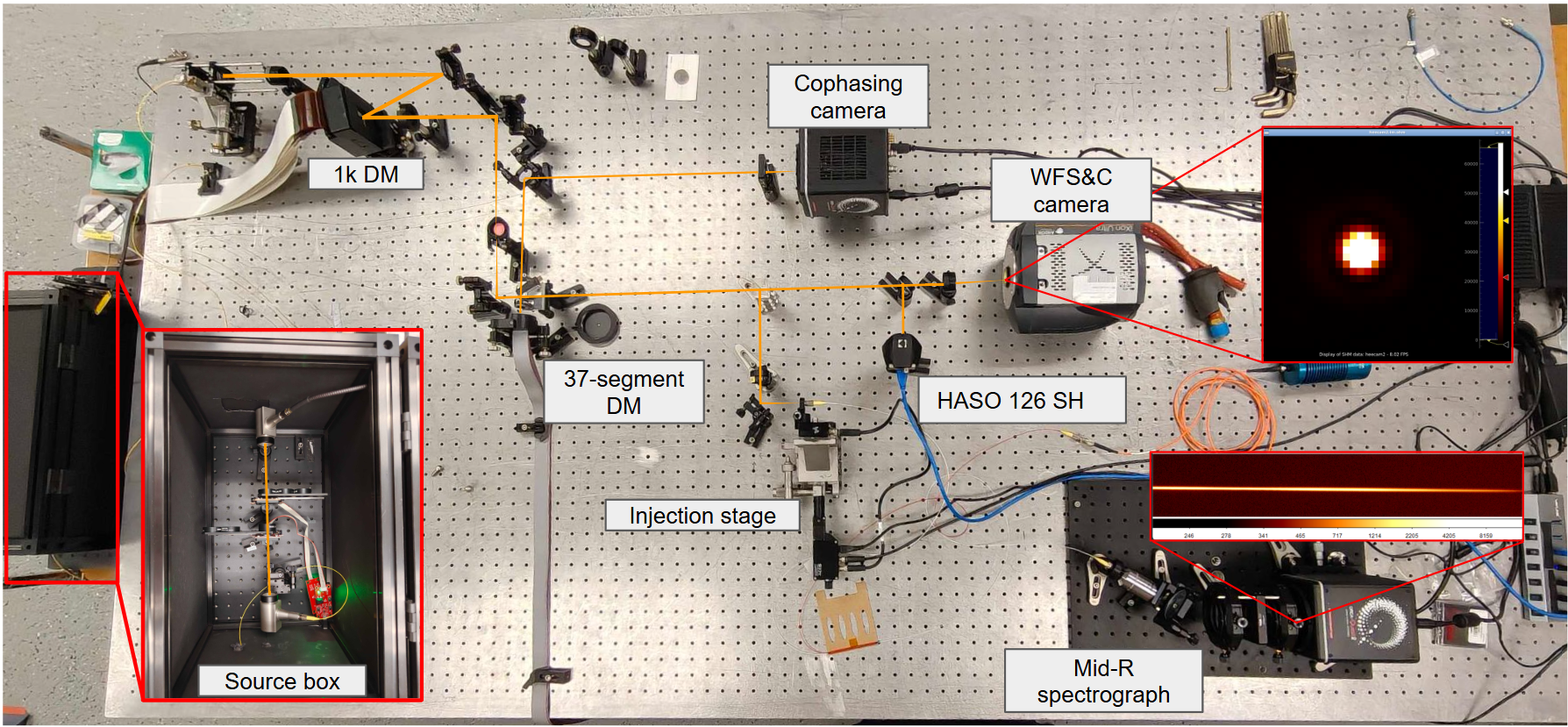}
    \caption{Physical implementation and commissioning status of the He`e-Lab optical bench. The orange beam trace highlights the path originating from the source box and propagating to the 1k DM. A polarizing beam splitter (PBS) divides the path: transmitted light feeds the primary WFS\&C and astrophotonic arms, while reflected light is routed through a double-pass QWP setup across the 37-segment DM (equipped with a hexagonal mask) before returning through the PBS to the cophasing camera. Insets display live focal plane telemetry (top right) and spectrograph output (bottom right).}
    \label{fig:bench_layout}
\end{figure}

The primary optical train of the He`e-Lab testbed has achieved complete physical alignment and initial commissioning, as illustrated in Figure~\ref{fig:bench_layout}. Light delivery originates from the enclosed supercontinuum source box and propagates sequentially through the primary active optics—the continuous BMC 1k DM and the 37-segment cophasing DM—before splitting into the diagnostic arm and astrophotonics injection module. Software control via \texttt{pyMilk} has been established across all active hardware, including DM grids, Zaber translation stages, and cameras.

Light routing in the cophasing arm is engineered for maximum photon efficiency. Following the 1k DM, a Polarizing Beam Splitter (PBS) transmits P-polarized light toward the WFS\&C and astrophotonic paths, while reflecting S-polarized light into the segmented mirror channel. This reflected beam passes through a Quarter-Wave Plate ($\text{QWP}$), strikes the 37-segment DM (fitted with an adjacent hexagonal pupil mask matching segment geometry), and reflects back through the $\text{QWP}$. This double-pass rotates the polarization state ($S \rightarrow P$), enabling full transmission back through the PBS cube directly to the cophasing camera without splitting intensity losses.

\begin{figure}[htbp]
    \centering
    \includegraphics[width=0.4\linewidth]{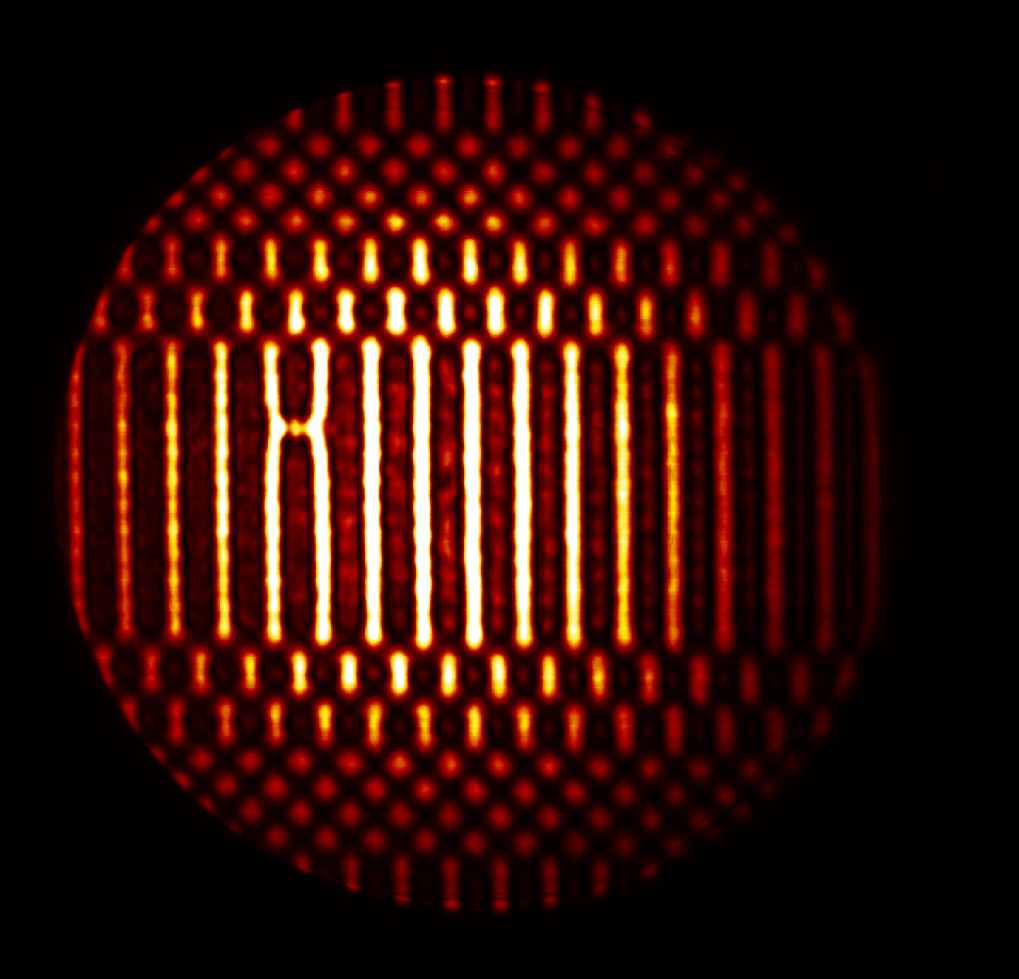}
    \caption{Direct pupil-plane image of the continuous DM surface subjected to high-spatial-frequency test patterns ($\pm$ voltage commands between neighboring actuators). The high-contrast intensity modulation confirms functional actuator-to-channel mapping, high dynamic range, and expected mechanical inter-actuator coupling.}
    \label{fig:dm_response}
\end{figure}

To confirm actuator command authority and channel-mapping integrity, dynamic test patterns were applied across the 1k DM grid. Figure~\ref{fig:dm_response} shows a re-imaged DM pupil plane during the application of an alternating high-frequency stroke pattern ($\pm$ displacement across adjacent actuators). The "high-contrast" grid image validates that the actuators are well-driven, responsive, and accurately mapped within the real-time software interface. We can also notice a couple of coupled actuators, as expected from the manufacturer documentation.

Diagnostic focus and beam stability are currently monitored using the focal-plane WFS\&C and cophasing cameras. Full integration of the HASO 126 Shack-Hartmann WFS is actively underway; current work focuses on migrating its control interface from a native Windows environment to a Linux system to enable direct shared-memory integration into the \texttt{CACAO}/\texttt{pyMilk} real-time pipeline.

\subsection{Astrophotonic Injection and SMF Coupling Optimization}
To achieve maximum throughput into Single-Mode Fibers (SMF), where spatial mode matching and sub-micron spatial positioning are critical, an automated optimization routine was developed within the software framework. Transverse ($XY$) positioning is managed by high-precision motorized Zaber actuators mounted on the 3-axis translation stage and driven via \texttt{pyMilk}, while focus ($Z$) is adjusted manually.

\begin{figure}[htbp]
    \centering
    \includegraphics[width=0.55\linewidth]{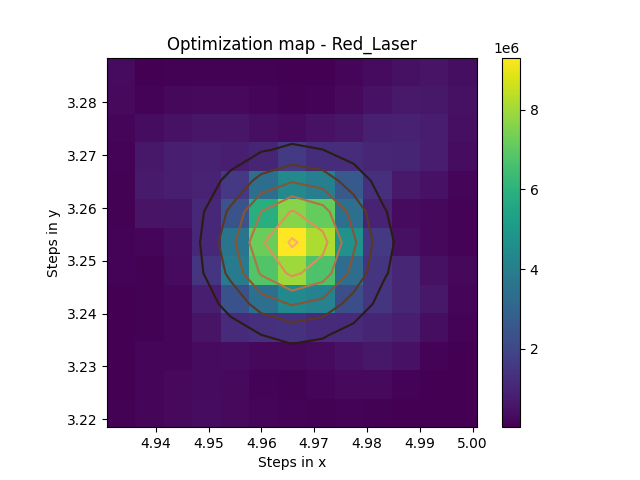}
    \caption{Empirical SMF coupling efficiency map acquired using a red laser source. The 2D raster scan maps transmitted intensity  across motorized $XY$ stage step coordinates, displaying smooth Gaussian contours centered on the fundamental fiber mode ($\text{LP}_{01}$).}
    \label{fig:coupling_map}
\end{figure}

The coupling optimization routine relies on a direct 2D spatial mapping and fitting approach:
\begin{enumerate}[nosep]
    \item \textbf{Transverse $XY$ Raster Scan:} The Zaber stages execute a 2D spatial grid scan across the fiber face while recording downstream flux at each step to construct an empirical coupling map (Figure~\ref{fig:coupling_map}).
    \item \textbf{2D Gaussian Surface Fitting:} Assuming the convolution of the incoming Point Spread Function (PSF) and the fundamental fiber mode ($\text{LP}_{01}$) is well-approximated by a 2D Gaussian profile, a non-linear least-squares Gaussian fit is applied to the spatial flux map. 
    \item \textbf{Centroid Extraction and Locking:} The fitted peak coordinates directly define the optimal injection position, enabling the Zaber stages to immediately command and lock onto the absolute maximum throughput coordinates without iterative gradient searches.
\end{enumerate}

This deterministic fitting routine provides a rapid and robust alignment baseline. In full operation, this stage positioning framework will operate in tandem with active phase profile adjustments on the upstream continuous 1k DM, enabling active correction of wavefront aberrators while holding optimal spatial fiber injection.

\section{FUTURE PROSPECTS AND STRATEGIC ROADMAP}
\label{sec:roadmap}
The commissioning of the He'e-Lab testbed establishes a baseline operational environment for sea-level technology validation. Moving forward, the facility will pursue several key technical milestones across its adaptive optics, segmented aperture control, and astrophotonics research tracks:
\begin{enumerate}
    \item \textbf{Closed-Loop AO Validation:} Following the Linux-native migration of the HASO WFS drivers, real-time closed-loop operations will be fully established within the \texttt{CACAO} software environment. This will enable high-speed wavefront correction using the BMC 1k DM to simulate and mitigate atmospheric turbulence.
    \item \textbf{Focal-Plane Wavefront Sensing for Cophasing:} The 37-segment hexagonal DM will serve as a dedicated testbed for non-common path aberration (NCPA) sensing and multi-segment cophasing. Advanced focal-plane WFS algorithms—such as Linearized Analytical Phase Diversity (LAPD~\cite{vievard_cophasing_2020}), Fast and Furious~\cite{bos_-sky_2020}, and DrWHO~\cite{Skaf2022} —will be deployed to achieve cophasing control.
    \item \textbf{Photonic Lantern Fed High-Resolution Spectrograph:} In collaboration with the Subaru/FIRST-PL team~\cite{vievard_spectroscopy_2024}, the injection arm will serve as the primary staging environment to develop, integrate, and validate a novel high-resolution spectroscopic module~\cite{hamner2026}. By unifying the software interfaces in the lab to match the SCExAO/FIRST-PL environment beforehand, the platform ensures rapid, full-system validation for seamless direct integration on-sky with minimal operational risk.
    \item \textbf{3D ULI Photonic Lantern Characterization:} As part of an undergraduate research initiative at UH Hilo, custom 3D photonic lanterns manufactured via Ultrafast Laser Inscription (ULI) will be integrated and characterized on the testbed.
    \item \textbf{NIR Arm Expansion:} To fully leverage the broadband output of the supercontinuum source ($500\,\text{nm}$--$2\,\mu\text{m}$), a dedicated Near-Infrared (NIR) arm operating from $\sim 900\,\text{nm}$ to $2\,\mu\text{m}$ is currently under development. This extension will expand the bench's spectroscopic and astrophotonic capabilities into the $Y$, $J$, and $H$ bands.
    \item \textbf{Intermediate development platform for Keck and Subaru:} We intend to implement Keck specific software structures, enabling wavefront control tests to be performed on the lab prior to being tested at the Keck telescope. Likewise, the current RTC on the He'e lab is the same as in SCExAO, making it an ideal platform for contributing to the technical collaboration between the two telescopes.   
    \item \textbf{Open-Access Collaborative Sandbox:} Designed as a highly accessible platform, He'e-Lab actively invites visiting researchers and external instrumentation teams to integrate custom experiments, validate photonic architectures, and test control algorithms in a low-risk environment.

\end{enumerate}

% Ultimately, mature hardware and software configurations validated inside the He`e-Lab sandbox will be prioritized for rapid on-sky translation to high-contrast instruments like SCExAO on the Subaru Telescope.

\section{CONCLUSION}
\label{sec:conclusion}

We have presented the architecture, hardware integration, software framework, and initial commissioning results for the Hawaii Experimental Engineering Lab (He'e-Lab) testbed. By pairing a high-order 1k-actuator continuous DM with a 37-segment PTT cophasing assembly, a HASO 126 Shack-Hartmann WFS, and a flexible astrophotonic injection stage, He'e-Lab provides a versatile, low-risk sea-level testbed for advanced wavefront control and photonic device validation/integration.

Initial bench commissioning verified complete optical alignment, direct low-level software control via \texttt{pyMilk} across active components, dynamic DM actuator responsiveness (with expected mechanical inter-actuator coupling), and automated single-mode fiber coupling via a deterministic 2D spatial raster scan and Gaussian fitting procedure. Full high-level loop control via the \texttt{CACAO} software environment is planned for near-term closed-loop testing.

As a sea-level sandbox mirroring the hardware and software environments of major observatories, He'e-Lab bridges the critical gap between bench-top concepts and facility-class deployment. Ongoing developments—including closed-loop AO integration, focal-plane cophasing algorithms, 3D ULI photonic lantern testing, and extension into the near-infrared—will further establish the facility as an accessible collaborative hub for technology development, pre-deployment risk mitigation, and student workforce training.

\acknowledgments 
The development of the CACAO software is supported by the National Science Foundation under award 2410616. The authors acknowledge the assistance of Gemini (Google) in manuscript preparation, specifically for assistance with LaTeX formatting and language polishing. NS acknowledges support from the Heising-Simons Foundation through the 51 Pegasi b Fellowship. JC acknowledges support from NSF REU Grant No. 2548017. 
The authors wish to recognize and acknowledge the very significant cultural role and reverence that the summit of Maunakea has always had within the indigenous Hawaiian community, and are most fortunate to have the opportunity to conduct observations from this mountain.

\bibliography{report} 
\bibliographystyle{spiebib} 
\end{document}